# Effect of chromium disorder on the thermoelectric properties of Layered-antiferromagnet $CuCrS_2$

**Girish Chandra Tewari\*, Tripurari Sharan Tripathi and Ashok Kumar Rastogi**
**School of Physical Sciences, Jawaharlal Nehru University, New Delhi -110067, India**

## Abstract

Layered-antiferromagnetic compound $CuCrS_2$ has been prepared by different methods. The analysis of X-ray diffraction patterns of different samples gave significant amount of vacancy-disorder of Cr-atoms within the layers. Extended period of sintering above $900^0C$ increases the transfer of Cr-atoms to the interstitial sites between the layers. This disorder has marginal effect on the Antiferromagnetic properties. The electrical conductivity is increased and the thermoelectric power remains positive and quite high between 150-400μV/K in the paramagnetic state around room temperature with increase in disorder in different samples. We interpret the temperature dependence of electrical resistivity and thermoelectric power due to the localization of carriers by interstitial defects and the formation of magnetic polarons in the paramagnetic phase of $CuCrS_2$.

## Introduction:

Ternary chromium chalcogenides ACrX$_2$ (A = Li, Na, K, Cu or Ag and X=S, Se) have been extensively investigated for their antiferromagnetic properties. Cr-atoms in these compounds form hexagonal layers and are separated by the layer of non-magnetic A-atoms. In CuCrS$_2$, Cr occupies octahedral sites while Cu occupies tetrahedral sites of alternate layers surrounded by closed packed sulfur or selenium atoms. In this family of compounds the competing interactions from the distant neighbors of the Cr-atoms cause low magnetic ordering temperatures T$_N$ lie between 20K-55K compared to the paramagnetic Curie temperatures of between -250K to +250K. In the ordered state the magnetic moments lie in the plane of chromium atoms and possess a non-collinear (Helical) spin structure [1, 2]. The magnetic ordering temperature T$_N$ is 40K for CuCrS$_2$ [1-5].

The nature of charge transport in these compounds is not well understood. Different studies conclude either insulating nature or narrow band gap magnetic semi-conduction [3, 4]. In our recent publication we have reported that the temperature dependence of electronic conductivity in CuCrS$_2$ is complex and sensitive to the native defects. The off-stoichiometry and sintering temperature have important effects on thermoelectric power and thermal conductivity of polycrystalline pellets. The electronic properties of CuCrS$_2$ were shown to be quite promising for the thermoelectric applications [5]. In the present study we report the effect of Cr-vacancy defects on the electronic transport properties of the layered antiferromagnet CuCrS$_2$. We have prepared polycrystalline samples of CuCrS$_2$ by different methods and treated them with different sintering conditions. We present the structural details by the analysis of x-ray diffraction patterns and the transport properties of three differently prepared samples of CuCrS$_2$.

**Preparation:**

The samples were prepared by two different methods. In the first method sample was prepared by reacting binary sulfides CuS and Cr$_2$S$_3$ in 2:1 molar ratio. The powders were mixed thoroughly and pelletized at 10-ton pressure. The reaction in a sealed quartz tube at 700$^0$C for five days gave CuCrS$_2$ and excess sulfur was collected at the colder end of the tube. We identify this phase, prepared at low temperature, as sample-1(700C). In the second method CuCrS$_2$ was prepared from the reaction of elements Cu(99.9%), Cr(99.9%) and S(99.9%) in 1:1:2 atomic ratios in a sealed quartz tube at 900$^0$C. The reacted mass was grounded thoroughly, pelletized and was sintered separately at a high temperature of 900$^0$C for 2-days and for 8-days. These samples are identified as 2a(900C) and 2b(900C) respectively. The sintering at higher temperature gave highly textured pellets for both 2a/2b(900C) -- with the growth of 00l-planes of hexagonal crystallites parallel to its surface as was easily inferred from the x-ray diffraction pattern obtained directly from the surface of the pellet (Figure 1a).

**Crystal Structural:**

**X-ray Diffraction:** In figure 1a we have presented the x-ray diffraction patterns of sample-1(700C) and 2a/2b(900C) prepared respectively from the binary sulfides and from the pure elements. The data were taken from a Philips Diffractometer by using Cu K$_\alpha$ radiation source. In case of 2a/2b(900C), the pattern was obtained from the x-ray diffraction directly from the surface of the respective pellets. The large increase of 00l reflections in these cases is due to the texturing of the pellets on sintering them at high temperatures.

**Refinement of structure:** We chose "Cosine Fourier series" function, among the functions included in the refinement program, for the background radiation and properly refined the background coefficients for our pattern. The lattice parameters, atomic positions and occupancy of chromium atoms on the two different sites were refined sequentially using GSAS-EXPGUI [6]. The chosen atomic positions for the crystal structure of CuCrS$_2$ are shown in figure 2. All the atoms occupy 3(a) 00z sites of space group R3m. We chose z=0.15 for Cu-atom and z=0.27 for S(I) and z=0.73 for S(II) as suggested by the Bonger et al. for the ideal structure and with the full occupancy of one for them [1]. While Cr-atoms in our case were distributed over the intra-layer sites z=0 and inter-layer sites within the Cu-layers with z = 0.50. We chose a combined occupancy of both for Cr-sites as one since it gave the best minimization of R(F$^2$) and $\chi^2$ parameters for all the three samples. This choice was also based on our results of the chemical analysis by SEM (EDAX) which gave the same relative concentration of Cu, Cr and S within 5% of the stoichiometric amount for all the samples prepared at low as well as at high temperatures. The fitting parameters for the x-ray patterns were not significantly affected by the 5% change in the atomic concentration. The alternative choice of disorder in the Cu occupancy, as used for the single crystal with composition Cu$_{1.03}$Cr$_{0.91}$S$_2$ by Le Nagard et al. was found to be unsuitable for the satisfactory refinement in case of our polycrystalline samples [3].

The results of refined parameters -- lattice constant, the z-values of occupied sites and the occupancy of Cr-sites along with the least square fit parameters $\chi^2$ and R (F$^2$) -- are presented in Table1. The required parameter for the preferred orientation corrections in the different patterns is also noted. The fitting of our x-ray patterns are quite satisfactory as can be seen from the reasonable value of R(F$^2$) and $\chi^2$ in all the three cases that compares well with the similar studies. Our analysis gives a significant amount of interstitial Cr(II) atoms in all the samples, with largest concentration of 20% for 2b(900C) phase that was sintered at 900$^0$C over extended period of 8 days. There is, however, no significant change in their unit cell parameters that were found to be similar to the single crystal reported by Le Nagard et al. [3].

**Transport Properties results and discussion:**

**Experimental details:** The electrical conductivity measurements between 15K and 300K were performed by four point probe method using silver paste for the electrical contacts. The thermopower of the pellets and the crystal flakes was studied by measuring Seebeck coefficient $S$ (= $\Delta V/\Delta T$) with respect to copper from 15K to 300K using a differential method of measurement. In this method, at a stabilized temperature a small temperature gradient is generated across the sample length and thermoelectric voltage is measured with respect to copper leads. The spurious and offset voltages in the circuit were eliminated by reversing the temperature gradient and averaging the voltages. The apparatus was tested for the accuracy by measuring $S$ on a thin piece of pure Lead with respect to copper at low temperatures. The accuracy of S is about 5%.

**Electrical Resistivity:** Electrical resistivity of $CuCrS_2$ at low temperatures, plotted as Log($\rho$) vs. T, is presented in the upper panel of figure3. In the lower panel of figure 3, we have shown the resistivity of a crystal flake and in its inset resistivity of a pellet up to 600K. All our polycrystalline samples show rise in resistivity on cooling towards the magnetic transition temperature, while resistivity of single crystal flakes show metal-like decrease on cooling with a shallow minimum below 50K. There is a sharp downward jump at $T_N$ in the resistivity of all the samples and it becomes almost temperature independent below $T_N$. This behavior reflects that electronic conduction in $CuCrS_2$ is affected by the magnetic order.

The effect of Cr-disorder on the resistivity can be clearly seen by comparing the behavior of 2b(900C) with the 1(700C) sample; the room temperature resistivity is considerably reduced by the increase in disorder, but in this case the rise on cooling is accompanied by a characteristic plateau in its resistivity between 100K to 200K. We believe that this behavior is the result of the interstitial occupancy of the Cr atoms quite similar to the $Cr_{1+x}S_2$ compounds. The later have similar layer structure with interlayer Cr-atoms and show antiferromagnetic transition between100K to 200K [7]. The metal–like dependence of the resistivity of single crystal flakes and the absence of the anomaly around 100-200K in this case may be due to absence of interstitial occupancy of Cr-atoms within the Cu-layers of $CuCrS_2$ as reported by Le Nagard et al. [3].

The overall resistivity behavior of the sintered pellets of $CuCrS_2$ at low temperatures is quite complex. The value of the resistivity of pellets and the crystal flakes are similar around room temperature. The single crystal flake above $T_N$ and polycrystalline pellet above room temperature (see inset of lower panel of figure 3) show metallic temperature dependence of resistivity. A Minimum in the resistivity of pellets is observed around room temperature which is similar to the crystal flakes where a minimum is observed below 50K. These observations clearly demonstrate that contrary to the previous reports the $CuCrS_2$ is not insulating, but show variable degree of localization of the conduction electrons [3, 4]. The large increase in the low temperature resistivity in case of the of polycrystalline pellets as compared to the resistivity measured in the plane of the layers in the crystal flakes may be related to the anisotropy of the scattering and in case of polycrystalline samples to the localization effects perpendicular to the layers by the Cr-disorder and at the grain boundaries.

**Thermoelectric Power:** The Seebeck coefficient S of polycrystalline samples and a crystal flake is shown in figure 4. S is positive for all the samples including the flake and has a value around room temperature between 150-450/μV/K. S is found to vary smoothly across the magnetic transition temperature $T_N$; it increases rapidly on heating above the magnetically ordered phase and nearly saturates in the paramagnetic phase at high temperatures. It is unusual to see that the increase in the inter-layer chromium atoms in case of 2b(900C) causes an increase in thermopower compared to 2a(900C) sample while resistivity is reduced. For the polycrystalline pellets a large and nearly saturated behavior of thermopower is found at high temperatures, where the conductivity showed strong localization of the carriers. We explain this by the hopping conduction of small concentration of "localized" charge carriers, possibly by the formation of magnetic polarons in the sintered pellets. At sufficiently high temperatures, where the quantum effects are not important, S for the hopping conduction is independent of temperature and can be given by a classical Heike's formula $S = \frac{K}{e}\ln\left(\frac{x}{1-x}\right)$ where $x = \frac{n}{N}$; $n$=No. of localized charge carriers and $N$=Total No. of hopping sites [8]. At low temperatures all the compounds show rapid decrease in S where reduced localization effects, as the magnetic order is set, cause the band-conduction of the polarons.

**Conclusion:** Refinement of the x-ray diffraction patterns reveal that the sintering above $900^0C$ for an extended period gives significant concentration of chromium in the inter-layer sites of the layered-$CuCrS_2$. The inter-layer chromium has only marginal effect on the antiferromagnetic ordering temperature at $T_N$~40K. In this study we find that the increased chromium disorder reduces the resistivity, but increases the localization of the charge carriers causes the rise in the resistivity on cooling towards $T_N$. The thermopower behavior is not much affected which remains quite high in the paramagnetic phase. We explain the rise in resistivity on cooling towards $T_N$ as due to interplay of localization of charge carriers by the defects and grain boundaries and the increased magnetic scattering. The saturating behavior of thermopower at high temperatures can be explained by the formation of magnetic polarons where the Heike's formula for S is applicable [8].

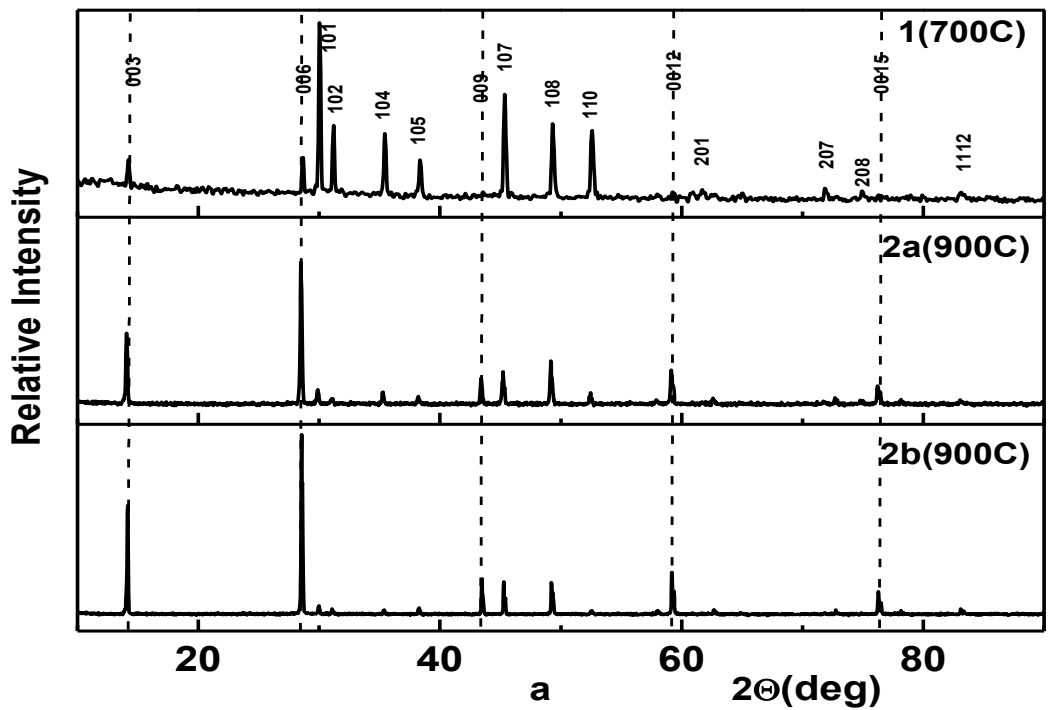

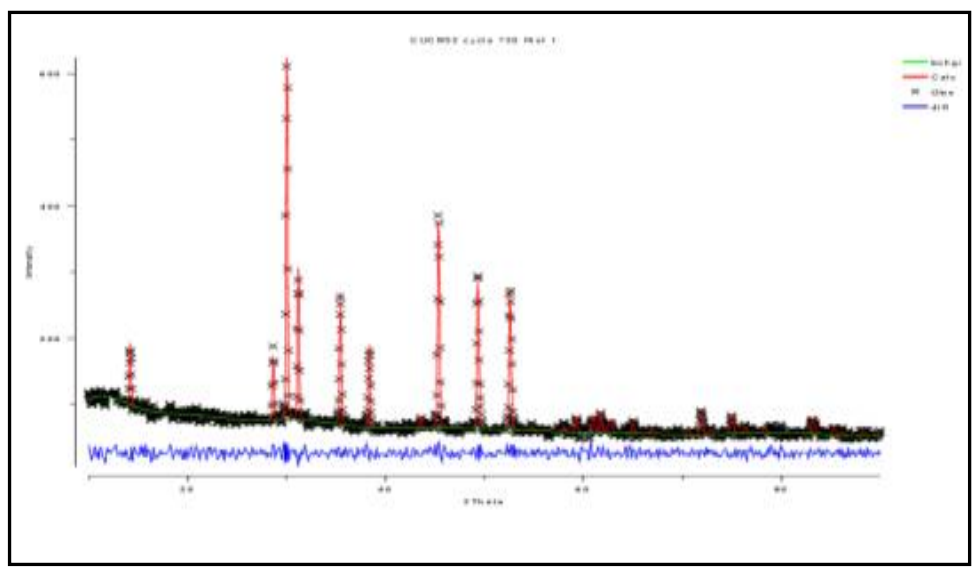

**Fig. 1. (a)** X-ray diffraction patterns of hexagonal-rhombohedral $CuCrS_2$ taken directly from the surface of polycrystalline pellets showing increased intensity of 00l reflections (marked by dashed line) in case of sintered samples 2a/2b(900C) samples prepared at high temperatures. **(b)** Calculated, observed and error patterns for the 1(700C) phase refined by GSAS (EXPGUI) program.

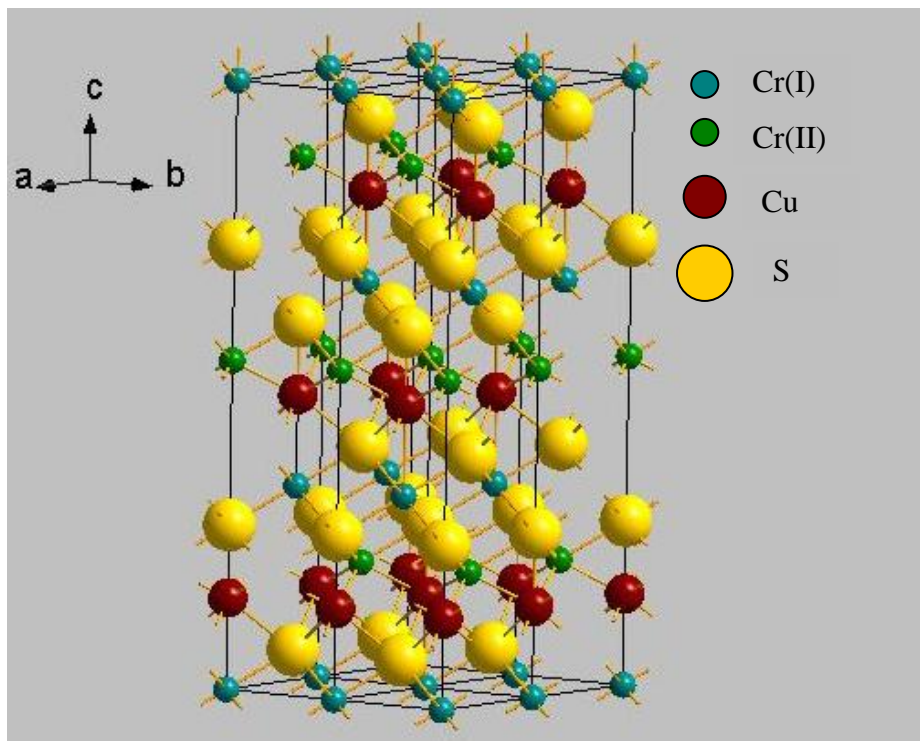

**Fig. 2.** Unit cell of CuCrS$_2$ showing layered structure. Cr(I) is the regular site while Cr(II) is inter-layer site occupied in disordered phases.

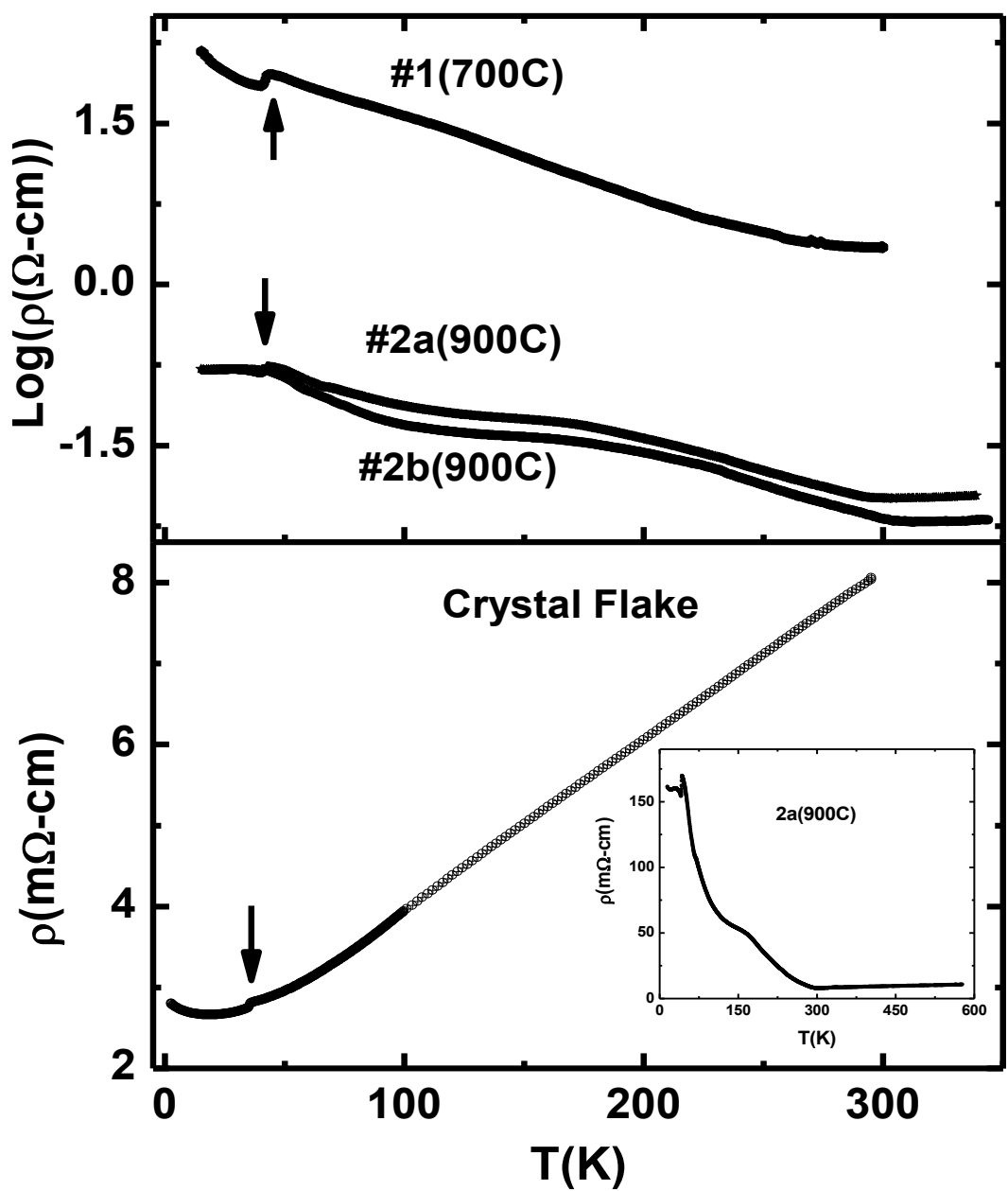

**Fig. 3.** Electrical resistivity Log(ρ) vs. T of CuCrS$_2$ pellets (upper panel) and crystal flake ρ vs. T (lower panel). Arrow marks the antiferromagnetic transition. A metal–like dependence above 300K and a plateau-like feature around 150K is shown for 2a (900) sample in the inset.

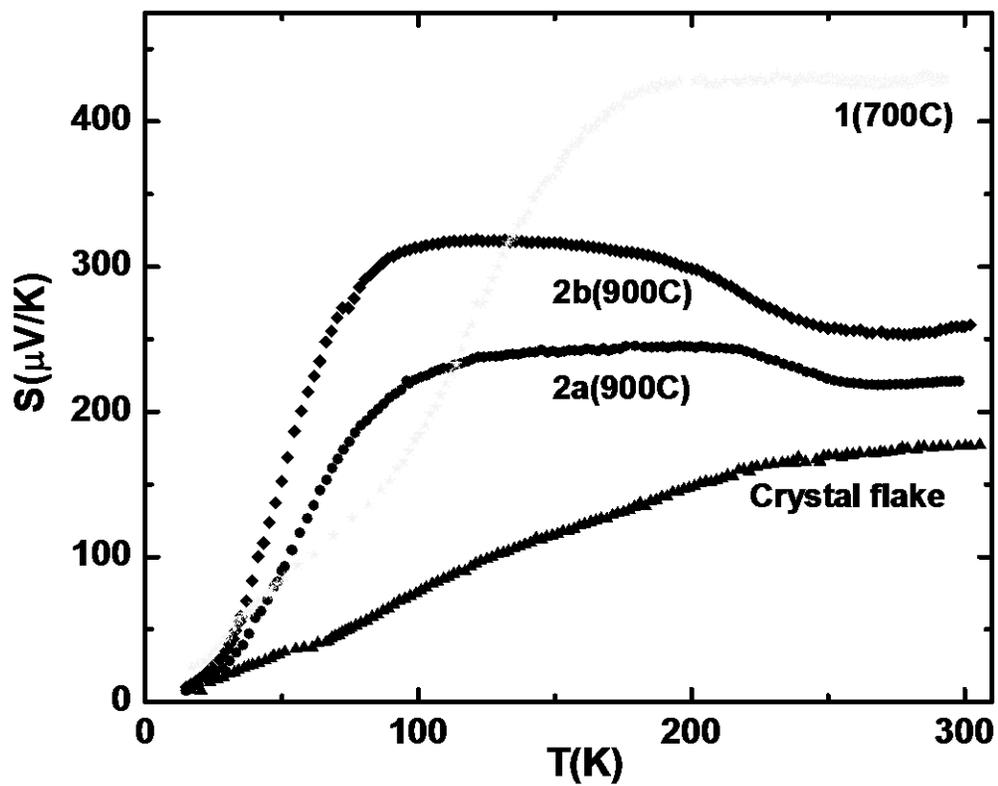

**Fig. 4.** Seebeck coefficients of different samples of $CuCrS_2$, showing rapid rise with temperature above the antiferromagnetic transition and saturation behavior at high temperatures.

Table 1: Refined structural parameters of $CuCrS_2$ (space group R3m) using GSAS (EXPGUI).

| | Sample-1(700$^0$C) | | Sample-2a/2b(900$^0$C) | | | |
| --- | --- | --- | --- | --- | --- | --- |
| | #1 | | #2a | | #2b | |
| Atom | z | Occupancy | z | Occupancy | z | Occupancy |
| Cr(I) | 0 | 0.95(1) | 0 | 0.84(1) | 0 | 0.81(1) |
| Cr(II) | 0.5230(5) | 0.05(1) | 0.5203(4) | 0.15(1) | 0.5230(1) | 0.19(1) |
| Cu | 0.1474(4) | 1.00 | 0.1487(1) | 1.00 | 0.1490(2) | 1.00 |
| S(I) | 0.2619(4) | 1.00 | 0.2640(3) | 1.00 | 0.2642(3) | 1.00 |
| S(II) | 0.7381(4) | 1.00 | 0.7359(6) | 1.00 | 0.7358(3) | 1.00 |
| a=3.4825(3), c=18.710(1) | | | a=3.4789(3), c=18.6861(8) | | a=3.4794(8), c=18.702(4) | |
| $\chi^2 = 0.316$, $R(F^2) = 0.10$ Preferred orientation coefficient(ratio) =0.931(6) | | | $\chi^2 = 1.49$, $R(F^2) = 0.08$ Preferred orientation Coefficient(ratio)=0.426(2) | | $\chi^2 = 1.80$, $R(F^2) = 0.08$ Preferred orientation coefficient(ratio) =0.370(3) | |

## *Acknowledgement:*

G. C. T and T. S. T acknowledge the Council of Scientific and Industrial Research (CSIR) India. We also acknowledge AIRF, JNU for X-ray diffraction measurement.

*E-mail: gctewari2002@gmail.com